\documentclass{evn2004}
\usepackage{txfonts}
\begin{document}
%
%%%% Personal Latex Definitions:-
\def\farcs{\hbox{$.\!\!^{\prime\prime}$}}
\def\kms      {\ifmmode {\rm km\,s}^{-1} \else km\,s$^{-1}$\fi}
\def\mujybm{${\rm \mu}$Jy\,beam$^{-1}$\fi}
\def\ltsim{\ifmmode\stackrel{<}{_{\sim}}\else$\stackrel{<}{_{\sim}}$\fi}
\def\gtsim{\ifmmode\stackrel{>}{_{\sim}}\else$\stackrel{>}{_{\sim}}$\fi}
%%%%%%%%%%%%%%%%%%%%%%%%%%%%%%%%%

   \title{Neutral hydrogen absorption at milliarcsecond resolutions:\\
The radio galaxy 3C\,293}

   \author{R. J. Beswick\inst{1}
          \and
          A. B. Peck\inst{2}
	\and
	G. B. Taylor\inst{3}
	\and
	G. Giovannini\inst{4}
	\and 
	A. Pedlar\inst{1}
          }

   \institute{Jodrell Bank Observatory, The University of Manchester,
Macclesfield, Cheshire, SK11 9DL, UK
         \and
             Harvard-Smithsonian Center for Astrophysics, SAO/SMA
Project, P.O. Box 824, Hilo, HI 9672, USA
	\and
	National Radio Astronomy Observatory, P.O. Box 1 Socorro, NM
87801, USA
	\and
	Istituto di Radioastronomia del CNR, via Gobetti 101, 40129
Bologna, Italy
             }

   \abstract{We present new milliarcsecond resolution observations of
the H{\sc i} absorption against the kiloparsec scale inner jet of the
radio galaxy 3C\,293. Using a combination of observations obtained
with global VLBI, MERLIN and the VLA we have imaged the strong and
extensive neutral hydrogen absorption against the radio core and jet
of this source across a wide range of angular scales.

In this proceedings we will present these new combined milliarcsecond
scale VLBI results alongside
our previous lower resolution MERLIN studies of the H{\sc i}
absorption in this source. This study will allow us to investigate the distribution and dynamics of the H{\sc i} absorption in
the centre of this source from scales of arcseconds to a few milliarcsecond.}

%You have 4 pages for an oral contribution, 
%and 2 pages for a poster contribution.  You have to upload the final 
%PostScript file {\bf before September 13th 2004} 
%in the ftp directory\\
%\centerline{{\tt ftp://ftp.oan.es/evn2004/}.}\\  
%After that,
%you should mail to {\tt evn2004@oan.es} 
%to give us notice about the submission of the paper.
%The book with the proceedings will be distributed during 
%the conference.

   \maketitle
%
%________________________________________________________________

\section{Introduction}

The study of the physical properties of the dust and gas at the
centres of active galaxies is of great observational interest. Its
significance is inextricably linked to the activity observed in these
sources since it is the gas and the dust that is ultimately the fuel
for the activity.   The dust
distribution can be studied using optical
and infrared imaging. The atomic gas can be studied using the 21-cm
line of neutral hydrogen (H{\sc i}). As well as giving clues to formation of
active galaxies, dust and gas may obscure the direct view to the
active galactic nuclei (AGN) and thus affect our ability to view the
central optical continuum and broad-emission-line region. 

With present day instruments sub-arcsecond
optical observations of the ionised gas and radio observations of the
synchrotron emission are relatively routine. There are relatively few methods by which the abundances
and kinematics of the components within neutral ISM can be studied in
galaxies with high angular resolution.  Components, such as dust, preclude
optical observations of the hearts of these sources and the present
array of mm-wavelength instruments do not currently provide the
angular resolution to study molecular gas at great detail.  Unfortunately atomic
hydrogen emission can only be observed by integrating over large
areas - typically tens of arcsec - because of its low ($\sim$100\,K)
brightness temperature and the current sensitivities of decimeter
arrays. However, provided the neutral gas is in front of a strong
source of radio continuum, absorption studies can be used to study the
atomic hydrogen on milliarcsecond scales. Atomic hydrogen absorption observations allow the derivation of parameters such as the transverse velocity gradients and structure of the neutral gas can then contribute to an understanding of the dynamics of the nuclear regions of active galaxies.

The radio galaxy 3C\,293 is a nearby (D=180\,Mpc) moderately large
two-sided Fanaroff-Riley type-II radio source with a disturbed optical
morphology and an unusually bright steep-spectrum extended core
component (Akujor et al 1996). 

The interstellar medium (ISM) of 3C\,293 is particularly rich. Strong detections of
molecular gas, for example CO 1$\rightarrow0$ observed in both
emission and absorption by Evans et al (1999) and extensive deep and
extremely broad H{\sc i} absorption detected against the kiloparsec
scale inner jet and core components (Haschick \& Baan 1985; Beswick,
Pedlar \& Holloway 2002; Morganti et al 2003). In this proceedings we
will present milliarcsecond observations using global VLBI, combined
with MERLIN and VLA, data to trace this deep absorption against the
inner jet of 3C\,293.

\begin{figure*}
   \centering
  \vspace{257pt}
\includegraphics{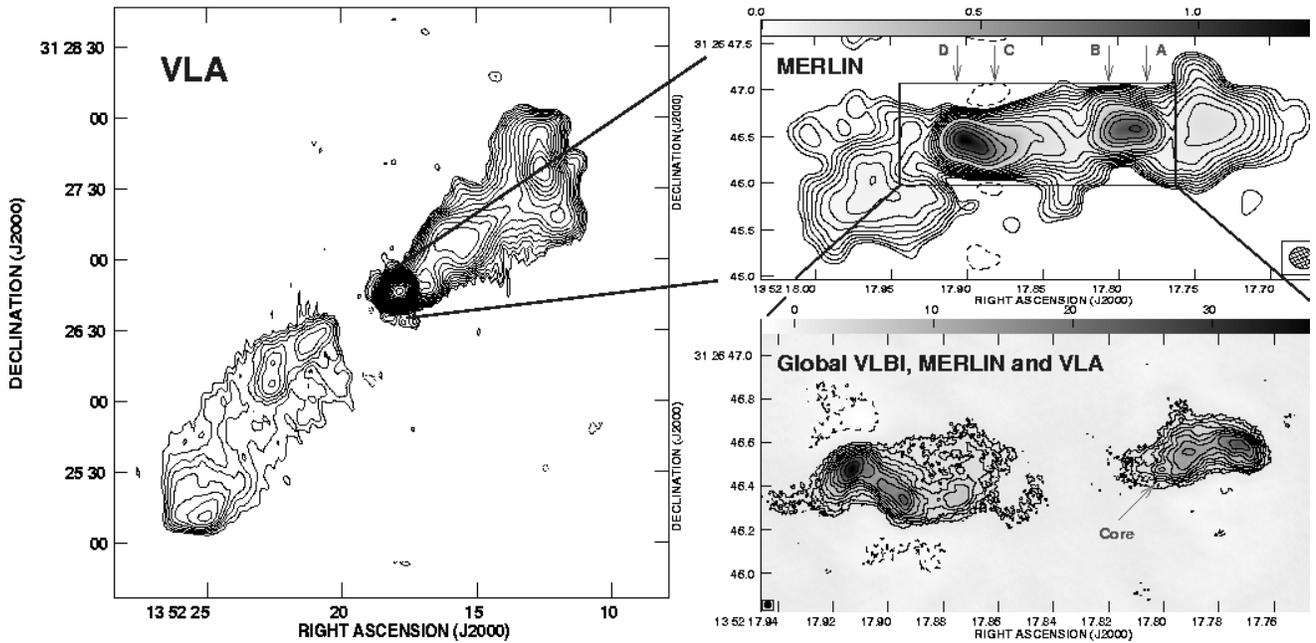}
  
 \caption{From kpc to pc: the multi-scale sub-arcsecond 1.359\,GHz radio continuum
structure of the inner jet in 3C\,293 as seen by the VLA, MERLIN and VLBI. The
right-hand upper panel shows a MERLIN image with a synthesised beam of
0\farcs23$\times$0\farcs20. This image is contoured at $\sqrt2$
multiples of 5mJy\,beam$^{-1}$. The right-hand lower panel shows the radio
structure with a 30\,mas angular resolution, derived from the
combination of VLA, including Pie Town, MERLIN and Global VLBI
observations. This image is contoured at multiples of $\sqrt2$ times
1.3\,mJy\,beam$^{-1}$. Whereas the left-hand image shows the large
scale radio structure of 3C\,293 as observed with the VLA. 
         \label{fig1}
         }
  \end{figure*}
\section{Observations \& Data Reduction}

3C\,293 was observed independently using the UK's Multi-Element Radio
Linked Interferometric Array (MERLIN) on April 8th 1998, a Global VLBI on
November 18th 1999 and using the VLA, in it's A-configurationand
including the VLBA antenna at Pie Town on December 15th 2000. All of
these observations were made at 1359\,MHz, the frequency of the
redshifted 21\,cm line H{\sc i} at 3C\,293.

Standard phase and amplitude calibration were applied to each of these
three data sets individually, using OQ208 as a phase calibration
source in all cases (as described in Beswick et al. 2002, 2004). The global VLBI data and VLA$+$PT data were averaged in frequency to
correspond with the lower velocity resolution MERLIN data, and the central
frequencies were shifted slightly so that the frequency range and channel
numbers in all data-sets matched exactly. The data-sets were then
concatenated by combining the VLA$+$PT and MERLIN data, and then the VLBI
data, with iterations of self-calibration at each stage.  The relative
weights for the data from each array were also checked during the process.
The combined data were subsequently Fourier transformed and deconvolved
using a circular 30\,mas restoring beam to form a 2048$\times2048\times$23
spectral line cube to which standard spectral line routines within {\sc
aips} were applied.

\section{Results \& Discussion}
\subsection{The radio continuum  jet structure}
 
In Fig.\,1 the radio structure of the jets of 3C\,293 are shown on
scales ranging from
several tens of kiloparsec to a few parsces (1 arcsec= 815\,pc).  As
can be seen on the MERLIN and VLBI scales the inner radio jet in
3C\,293 follows an almost east-west orientation. The orientation of
these inner hotspots and jet are misaligned by $\sim 45\deg$ to the
large-scale FR-II radio lobes (Fig.\,1 left hand panel; Bridle et al
1981; Beswick et al 2004). In the 0\farcs2 angular resolution MERLIN
1.3\,GHz observations, the inner jet of 3C\,293 shows at least four
distinct radio components (labelled A, B, C \& D in Fig.\,1a) with two
more diffuse lobe-like components extending on either side of the
central region. At this resolution and frequency it is unclear which
radio continuum component is coincident with the AGN. However, due 3C\,293's
 steep spectrum $\alpha\approx -1$ (S$_{\nu}\propto\nu^{-\alpha}$) (Akujor et
al 1996), the position of the core becomes apparent at higher frequencies.

On the largest observed scales (Fig.\,1) the $\sim 10$ kiloparsec
scale jet has a P.A.$\sim45\degr$. This jet trajectory is significantly
discordant with that of the inner jet structure implies a large
position angle change of the jet emission during the lifetime of the
radio source. This position angle change is indicative of the radio
emission having undergone multiple (at least two) phases of jet
emission, implying that the inner radio emission observed at
sub-arcsecond angular resolutions is a younger outburst. In this
scenario the large scale jet emission has arissen from an outburst
$\sim10^5$ years old. 

\begin{figure*}
   \centering
   \vspace{407pt}
   \includegraphics{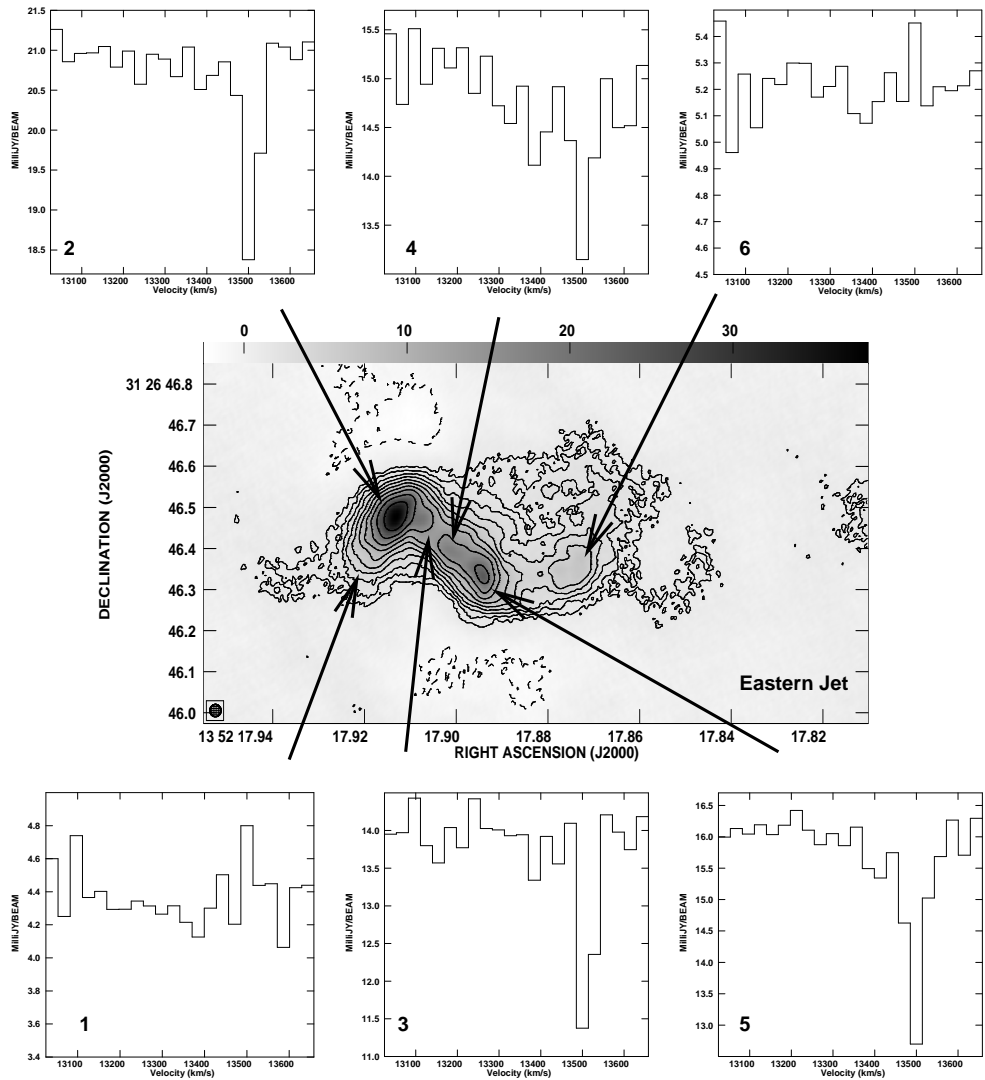}
      \caption{A montage of the eastern half of the inner radio jet
along with seletced H{\sc i} absorption spectra. 
         \label{fig2}
         }
   \end{figure*}

\subsection{The sub-kiloparsec scale absorption structure}

In earlier studies of the H{\sc i} absorption against 3C\,293 using WSRT (Shostak et
al 1983), the VLA (Haschick \& Baan 1985) and MERLIN (Beswick et al
2002), deep, multi-component H{\sc i} absorption was detected against
inner jet region. More recent high sensitivity, broad
bandwidth observations using WSRT by Morganti et al (2003) have also
shown that this nuclear H{\sc i} absorption possess an extremely broad
blueshifted wing of $\sim$1000\,km\,s$^{-1}$. In this study (see
Beswick et al 2004) we investigate further the deep and `relatively'
narrow H{\sc i} absorption against the inner jet with milliarcsecond
angular resolution. 

Beswick et al (2002) imaged the H{\sc i} absorption against the inner
jet of 3C\,293 (see Fig.\,1 top right) at $\sim$0\farcs2 with
MERLIN. At these angular resolutions it possible to separate two
absorbing components. Against the eastern half of the inner jet
structure the absorption line-widths are $\sim$40\,km\,s$^{-1}$
compared to the considerably broader line structure observed against
the western portion of the source. Consequently in the following to
sections we will discuss each of these spatially resolved components separately.  
\begin{figure*}
   \centering
   \vspace{407pt}
   \includegraphics{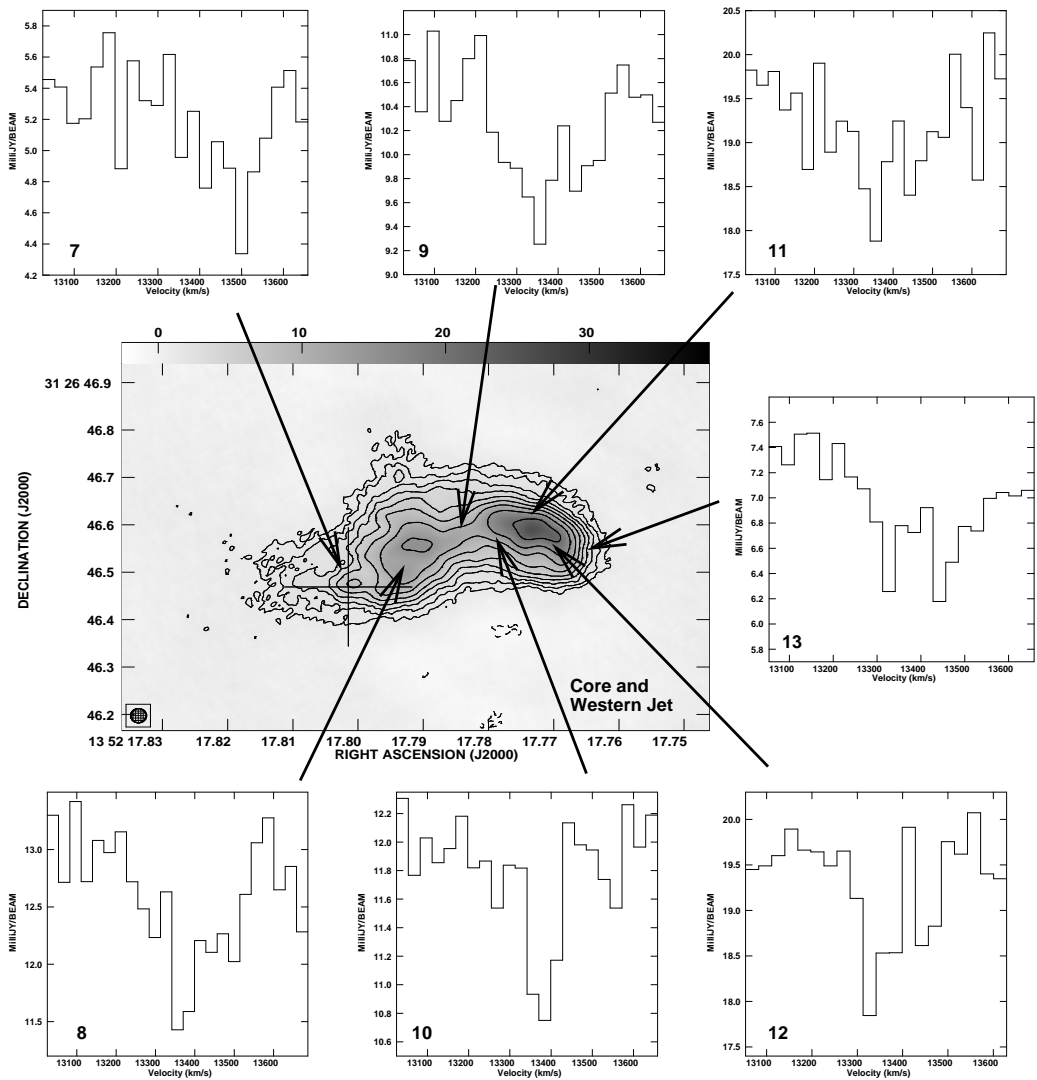}
      \caption{A montage of the western half of the inner radio jet
along with seletced H{\sc i} absorption spectra. The position of the
core, as measured at 5\,GHz with MERLIN by Akujor et al (1996) is
marked by a cross.
         \label{fig2}
         }
   \end{figure*}

\subsubsection{Against the eastern inner jet:- Narrow H{\sc
i} absorption}

As can be seen in Fig.\,2 the absorption against the eastern jet
components has narrow line-widths ($\sim$40\,km\,s$^{-1}$) and is
centred approximately on the rest velocity of 3C\,293
($\sim$13500\,km\,s$^{-1}$). In general this narrow absorption is
often attributed to more ambient, less disturbed foreground gas which
is lying at some distance from the centre of a galaxy. In the case of
3C\,293 this seems the most likely explanation for this narrow
component. However with these new high resolution observations its is
clear that this gas is not a homogeneous screen but shows distinct
spatial correlation with the positions of a cross nuclear dust lane
seen in {\it HST} observations (Beswick et al 2002, 2004). The
correlation of the lines of sight of these two components (H{\sc i}
and dust) suggests that both are probably physically associated. In
addition these new high resolution data, resolve a shallow velocity
gradient of $\sim$46\,km\,s$^{-1}$\,arcsec$^{-1}$ along a
P.A.$\sim60\rightarrow65\degr$. This is consistent with a velocity
gradient of ionised gas traced out to a radius of $\sim$10\arcsec ($\sim$8kpc)
using long-slit spectroscopy (van Breugel et al. 1984).  Consequently
it can be reasonable concluded that the dust, narrow H{\sc i}
component and the ionised gas are associated, situated at similar radii
away from the centre of the galaxy and are most likely undergoing
galactic rotation. 

\subsubsection{Against the western inner jet and core:- Broad H{\sc
i} absorption}

Against the western portion of the inner radio jet and the region
associated with the core broad, multi-component H{\sc i} absorption is
detected (see Fig.\,3). In the close vicinity of the core and western
jet region the absorption predominately displays two velocity
components.  Although, especially in these VLBI data, much of the
illuminating background radio continuum is resolved away, this complex
absorption does show some velocity structure on linear scales ranging
from $\sim$0.5\,kpc to $\sim$50\,pc. These absorbing structures can be
interpreted in two ways either as, consistent with two unrelated H{\sc
i} gas systems along the line of sight to the core, or with a steep
velocity ($\sim$410\,km\,s$^{-1}\approx$0.34\,km\,s$^{-1}$\,pc$^{-1}$)
centred upon the core. If the latter is assumed to be correct then the
mass enclosed by this rotating gas ring would be at least
1.7$\times$10$^9$\,M$_{\odot}$ within 400\,pc of the core.

%__________________________________________________ One column table
%   \begin{table}
%      \caption[]{H{\sc i} absorption properties.}
%         \label{KapSou}
%     $$ 
%         \begin{array}{p{0.5\linewidth}l}
%            \hline
%            \noalign{\smallskip}
%            Source      &  T / {[\mathrm{K}]} \\
%            \noalign{\smallskip}
%            \hline
%            \noalign{\smallskip}
%            Yorke 1979, Yorke 1980a & \leq 1700^{\mathrm{a}}     \\
%           Yorke 1979, Yorke 1980a & \leq 1700             \\
%            Kr\"ugel 1971           & 1700 \leq T \leq 5000 \\
%            Cox \& Stewart 1969     & 5000 \leq             \\
%            \noalign{\smallskip}
%            \hline
%         \end{array}
%     $$ 
%\begin{list}{}{}
%\item[$^{\mathrm{a}}$] This is footnote a
%\end{list}
%   \end{table}
%
%
%                                                One column figure
%----------------------------------------------------------- S_vib
   
%
%______________________________________________________________

\begin{acknowledgements}
RJB acknowledges PPARC support. The European VLBI Network is a joint facility of European, Chinese, 
South African and other radio astronomy institutes funded by their 
national research councils. The VLA and VLBA are operated by the
National Radio Astronomy Observatory (NRAO). NRAO is a facility of the
National Science Foundation operated under cooperative agreement by
Associated Universities, Inc. MERLIN is a national facility operated
by The University of Manchester on behalf of PPARC in the UK.

\end{acknowledgements}

\end{document}